%Paper: hep-th/9408036
%From: ANNA CERESOLE 3911564 7358 <CERESOLE@polito.it>
%Date: Fri, 5 Aug 1994 20:19:56 GMT+1

%%%%%%%%%%%%%%%%%%%%%%%%%%%%%%%%%%%%%%%%%%%%%%%%%%%%%%%%%%%%%%%%%%%%%%
\input harvmac
%%%%%%%%%%%%%%%%%%%%%%%%%%%%%%%%%%%%%%%%%%%%%%%%%%%%%%%%%%%%%
\newif\ifdraft

\noblackbox
\catcode`\@=11
\newif\iffrontpage
%%%%%%%%%%%%%%%%%%%%%%%%%%%%%%%%%%%%%%%%%%%%%%%%%%%%%%%%%%%%%%
%%%%% figures
%%%%%%%%%%%%%%%%%%%%%%%%%%%%%%%%%%%%%%%%%%%%%%%%%%%%%%%%%%%%%%
\def\figin{\epsfcheck\figin}\def\figins{\epsfcheck\figins}
\def\epsfcheck{\ifx\epsfbox\UnDeFiNeD
\message{(NO epsf.tex, FIGURES WILL BE IGNORED)}
\gdef\figin##1{\vskip2in}\gdef\figins##1{\hskip.5in}%
\else\message{(FIGURES WILL BE INCLUDED)}%
\gdef\figin##1{##1}\gdef\figins##1{##1}\fi}
\def\DefWarn#1{}
\def\figinsert{\goodbreak\midinsert}
\def\ifig#1#2#3{\DefWarn#1\xdef#1{fig.~\the\figno}
\writedef{#1\leftbracket fig.\noexpand~\the\figno}%
\figinsert\figin{\centerline{#3}}\medskip%
\centerline{\vbox{\baselineskip12pt
\advance\hsize by -1truein\noindent\footnotefont%
\centerline{{\bf Fig.~\the\figno}~#2}}
}\bigskip\endinsert\global\advance\figno by1}
%%%%%%%%%%%%%%%%%%%%%%%%%%%%%%%%%%%%%%%%%%%%%%%%%%%%%%%%%%%%%%
%%%%% sizes, offsets etc
%%%%%%%%%%%%%%%%%%%%%%%%%%%%%%%%%%%%%%%%%%%%%%%%%%%%%%%%%%%%%%
\ifx\answ\bigans
\def\titleft{\titsm}
\magnification=1200\baselineskip=14pt plus 2pt minus 1pt
%
%%%%% unreduced mode: %%%%
%\voffset=0.35truein\hoffset=0.250truein
\advance\hoffset by-0.075truein
\hsize=6.15truein\vsize=600.truept\hsbody=\hsize\hstitle=\hsize
\else\let\lr=L
\def\titleft{\titla}
\magnification=1000\baselineskip=14pt plus 2pt minus 1pt
%
%%%%% reduced mode: %%%%%%%
%\hoffset=-0.5truein\voffset=-.1truein
\hoffset=-.48truein\voffset=-.1truein
\vsize=6.5truein
\hstitle=8.truein\hsbody=4.75truein
\fullhsize=10truein\hsize=\hsbody
\fi
%
%\hstitle=8.truein\hsbody=4.5truein
%\fullhsize=9.75truein\hsize=\hsbody
\parskip=4pt plus 15pt minus 1pt
%%%%%%%%%%%%%%%%%%%%%%%%%%%%%%%%%%%%%%%%%%%%%%%%%%%%%%%%%%%%%%
%%%%%  fonts
%%%%%%%%%%%%%%%%%%%%%%%%%%%%%%%%%%%%%%%%%%%%%%%%%%%%%%%%%%%%%%
%%%%%%%%%%%%%%%%%%%%%%%%%%%%%%%%%%%%%%%%%%%%%%%%%%%%%%%%%%%%%%

\font\titla=cmr10 scaled\magstep3
\font\tenmss=cmss10
\font\absmss=cmss10 scaled\magstep1

\font\twelvebf=cmbx10 scaled\magstep1

\newfam\mssfam
\font\footrm=cmr8  \font\footrms=cmr5
\font\footrmss=cmr5   \font\footi=cmmi8
\font\footis=cmmi5   \font\footiss=cmmi5
\font\footsy=cmsy8   \font\footsys=cmsy5
\font\footsyss=cmsy5   \font\footbf=cmbx8
\font\footmss=cmss8
\def\footfont{\def\rm{\fam0\footrm}
\textfont0=\footrm \scriptfont0=\footrms
\scriptscriptfont0=\footrmss
\textfont1=\footi \scriptfont1=\footis
\scriptscriptfont1=\footiss
\textfont2=\footsy \scriptfont2=\footsys
\scriptscriptfont2=\footsyss
\textfont\itfam=\footi \def\it{\fam\itfam\footi}
\textfont\mssfam=\footmss \def\mss{\fam\mssfam\footmss}
\textfont\bffam=\footbf \def\bf{\fam\bffam\footbf} \rm}
\def\tenpoint{\def\rm{\fam0\tenrm}
\textfont0=\tenrm \scriptfont0=\sevenrm
\scriptscriptfont0=\fiverm
\textfont1=\teni  \scriptfont1=\seveni
\scriptscriptfont1=\fivei
\textfont2=\tensy \scriptfont2=\sevensy
\scriptscriptfont2=\fivesy
\textfont\itfam=\tenit \def\it{\fam\itfam\tenit}
\textfont\mssfam=\tenmss \def\mss{\fam\mssfam\tenmss}
\textfont\bffam=\tenbf \def\bf{\fam\bffam\tenbf} \rm}
\ifx\answ\bigans\def\abstractfont{\tenpoint}\else
\def\abstractfont{\def\rm{\fam0\absrm}
\textfont0=\absrm \scriptfont0=\absrms
\scriptscriptfont0=\absrmss
\textfont1=\absi \scriptfont1=\absis
\scriptscriptfont1=\absiss
\textfont2=\abssy \scriptfont2=\abssys
\scriptscriptfont2=\abssyss
\textfont\itfam=\bigit \def\it{\fam\itfam\bigit}
\textfont\mssfam=\absmss \def\mss{\fam\mssfam\absmss}
\textfont\bffam=\absbf \def\bf{\fam\bffam\absbf}\rm}\fi
%
%%%%%%%%%%%%%%%%%%%%%%%%%%%%%%%%%%%%%%%%%%%%%%%%%%%%%%%%%%%%%%
%%%%% footnotes   (adapted from PHYZZX)
%%%%%%%%%%%%%%%%%%%%%%%%%%%%%%%%%%%%%%%%%%%%%%%%%%%%%%%%%%%%%%
\def\f@@t{\baselineskip10pt\lineskip0pt\lineskiplimit0pt
\bgroup\aftergroup\@foot\let\next}
\setbox\strutbox=\hbox{\vrule height 8.pt depth 3.5pt width\z@}
\def\vfootnote#1{\insert\footins\bgroup
\baselineskip10pt\footfont
\interlinepenalty=\interfootnotelinepenalty
\floatingpenalty=20000
\splittopskip=\ht\strutbox \boxmaxdepth=\dp\strutbox
\leftskip=24pt \rightskip=\z@skip
\parindent=12pt \parfillskip=0pt plus 1fil
\spaceskip=\z@skip \xspaceskip=\z@skip
\Textindent{$#1$}\footstrut\futurelet\next\fo@t}
\def\Textindent#1{\noindent\llap{#1\enspace}\ignorespaces}
\def\footnote#1{\attach{#1}\vfootnote{#1}}%

\def\foot{\attach\footsymbolgen\vfootnote{\footsymbol}}
\let\footsymbol=\star
\newcount\lastf@@t           \lastf@@t=-1
\newcount\footsymbolcount    \footsymbolcount=0
\def\footsymbolgen{\relax\footsym
\global\lastf@@t=\pageno\footsymbol}
\def\footsym{\ifnum\footsymbolcount<0
\global\footsymbolcount=0\fi
{\iffrontpage \else \advance\lastf@@t by 1 \fi
\ifnum\lastf@@t<\pageno \global\footsymbolcount=0
\else \global\advance\footsymbolcount by 1 \fi }
\ifcase\footsymbolcount \fd@f\star\or
\fd@f\dagger\or \fd@f\ast\or
\fd@f\ddagger\or \fd@f\natural\or
\fd@f\diamond\or \fd@f\bullet\or
\fd@f\nabla\else \fd@f\dagger
\global\footsymbolcount=0 \fi }
\def\fd@f#1{\xdef\footsymbol{#1}}
\def\space@ver#1{\let\@sf=\empty \ifmmode #1\else \ifhmode
\edef\@sf{\spacefactor=\the\spacefactor}
\unskip${}#1$\relax\fi\fi}
\def\attach#1{\space@ver{\strut^{\mkern 2mu #1}}\@sf}
%
%%%%%%%%%%%%%%%%%%%%%%%%%%%%%%%%%%%%%%%%%%%%%%%%%%%%%%%%%%%%%%
%%%%% References
%%%%%%%%%%%%%%%%%%%%%%%%%%%%%%%%%%%%%%%%%%%%%%%%%%%%%%%%%%%%%%
\newif\ifnref
\def\rrr#1#2{\relax\ifnref\nref#1{#2}\else\ref#1{#2}\fi}
\def\ldf#1#2{\begingroup\obeylines
\gdef#1{\rrr{#1}{#2}}\endgroup\unskip}

\def\doubref#1#2{\refs{{#1},{#2}}}

\nreffalse
\def\refout{\listrefs}
%
%%%%%%%%%%%%%%%%%%%%%%%%%%%%%%%%%%%%%%%%%%%%%%%%%%%%%%%%%%%%%%
%%%%%%% eq numbering
%%%%%%%%%%%%%%%%%%%%%%%%%%%%%%%%%%%%%%%%%%%%%%%%%%%%%%%%%%%%%%
\def\eqn#1{\xdef #1{(\secsym\the\meqno)}
\writedef{#1\leftbracket#1}%
\global\advance\meqno by1\eqno#1\eqlabeL#1}
\def\eqnalign#1{\xdef #1{(\secsym\the\meqno)}
\writedef{#1\leftbracket#1}%
\global\advance\meqno by1#1\eqlabeL{#1}}
%
%%%%%%%%%%%%%%%%%%%%%%%%%%%%%%%%%%%%%%%%%%%%%%%%%%%%%%%%%%%%%%
%%%%%%  macros for titlepage, marginnotes, etc
%%%%%%%%%%%%%%%%%%%%%%%%%%%%%%%%%%%%%%%%%%%%%%%%%%%%%%%%%%%%%%
\def\chap#1{\global\advance\secno by1\message{(\the\secno\ #1)}
%\ifx\answ\bigans \vfill\eject \else \bigbreak\bigskip \fi  %if
% desired
\global\subsecno=0\eqnres@t\noindent{\twelvebf\the\secno\ #1}
\writetoca{{\secsym} {#1}}\par\nobreak\medskip\nobreak}
%% FOLLOWING LINE CANNOT BE BROKEN BEFORE 70 CHAR
%% FOLLOWING LINE CANNOT BE BROKEN BEFORE 70 CHAR
%% FOLLOWING LINE CANNOT BE BROKEN BEFORE 70 CHAR
\def\eqnres@t{\xdef\secsym{\the\secno.}\global\meqno=1\bigbreak\bigskip}
\def\sequentialequations{\def\eqnres@t{\bigbreak}}\xdef\secsym{}
\global\newcount\subsecno \global\subsecno=0
\def\sect#1{\global\advance\subsecno
by1\message{(\secsym\the\subsecno. #1)}
\ifnum\lastpenalty>9000\else\bigbreak\fi
\noindent{\bf\secsym\the\subsecno\ #1}\writetoca{\string\quad
{\secsym\the\subsecno.} {#1}}\par\nobreak\medskip\nobreak}
%%%%%%%%%%%%%%%%%%%%%%%%%%%%%%%%%%%%%%%%%%%%%%%%%%%%%%%%%%%%%%
%\def\chap#1{\newsec{#1}}
\def\chapter#1{\chap{#1}}
\def\section#1{\sect{#1}}
\def\\{\ifnum\lastpenalty=-10000\relax
\else\hfil\penalty-10000\fi\ignorespaces}
\def\note#1{\leavevmode%
\edef\@@marginsf{\spacefactor=\the\spacefactor\relax}%
\ifdraft\strut\vadjust{%
\hbox to0pt{\hskip\hsize%
\ifx\answ\bigans\hskip.1in\else\hskip .1in\fi%
\vbox to0pt{\vskip-\dp
%\vskip4pt
\strutbox\sevenbf\baselineskip=8pt plus 1pt minus 1pt%
\ifx\answ\bigans\hsize=.7in\else\hsize=.35in\fi%
\tolerance=5000 \hbadness=5000%
\leftskip=0pt \rightskip=0pt \everypar={}%
\raggedright\parskip=0pt \parindent=0pt%
\vskip-\ht\strutbox\noindent\strut#1\par%
\vss}\hss}}\fi\@@marginsf\kern-.01cm}
\def\titlepage{%
\frontpagetrue\nopagenumbers\abstractfont%
\hsize=\hstitle\rightline{\vbox{\baselineskip=10pt%
{\abstractfont\pubnum}}}\pageno=0}
\frontpagefalse
\def\pubnum{}
\def\pdate{\number\month/\number\yearltd}
\def\makefootline{\iffrontpage\vskip .27truein
\line{\the\footline}
%\vskip -.1truein\line{\pdate\hfil}
\vskip -.1truein\leftline{\vbox{\baselineskip=10pt%
{\abstractfont\pdate}}}
\else\vskip.5cm\line{\hss \tenrm $-$ \folio\ $-$ \hss}\fi}
\def\title#1{\vskip .7truecm\titlestyle{\titleft #1}}
\def\titlestyle#1{\par\begingroup \interlinepenalty=9999
\leftskip=0.02\hsize plus 0.23\hsize minus 0.02\hsize
\rightskip=\leftskip \parfillskip=0pt
\hyphenpenalty=9000 \exhyphenpenalty=9000
\tolerance=9999 \pretolerance=9000
\spaceskip=0.333em \xspaceskip=0.5em
\noindent #1\par\endgroup }
\def\autskip{\ifx\answ\bigans\vskip.5truecm\else\vskip.1cm\fi}
\def\author#1{\vskip .7in \centerline{#1}}

\def\address#1{\ifx\answ\bigans\vskip.2truecm
\else\vskip.1cm\fi{\it \centerline{#1}}}
\def\abstract#1{
\vskip .5in\vfil\centerline
{\bf Abstract}\penalty1000
{{\smallskip\ifx\answ\bigans\leftskip 2pc \rightskip 2pc
\else\leftskip 5pc \rightskip 5pc\fi
\noindent\abstractfont \baselineskip=12pt
{#1} \smallskip}}
\penalty-1000}
\def\endpage{\tenpoint\supereject\global\hsize=\hsbody%
\frontpagefalse\footline={\hss\tenrm\folio\hss}}
\def\ack{\goodbreak\vskip2.cm\centerline{{\bf Acknowledgements}}}
%%%%%%%%%%%%%%%%%%%%%%%%%%%%%%%%%%%%%%%%%%%%%%%%%%%%%%%%%%%%%%
\def\a{\alpha} \def\b{\beta} \def\d{\delta}
\def\e{\epsilon} \def\c{\gamma}
\def\G{\Gamma} 
 
\def\cA{{\cal A}} 
 \def\cD{{\cal D}}

 \def\cV{{\cal V}}
\def\cW{{\cal W}}
\def\IGa{\ralax{{\rm I}\kern-.18em \Gamma}}
\def\IZ{{\hbox{{\rm Z}\kern-.4em\hbox{\rm Z}}}}
\def\IR{{\hbox{{\rm I}\kern-.4em\hbox{\rm R}}}}
\def\nup#1({Nucl.\ Phys.\ $\us {B#1}$\ (}
\def\plt#1({Phys.\ Lett.\ $\us  {#1}$\ (}
\def\cmp#1({Comm.\ Math.\ Phys.\ $\us  {#1}$\ (}
\def\prp#1({Phys.\ Rep.\ $\us  {#1}$\ (}
\def\prl#1({Phys.\ Rev.\ Lett.\ $\us  {#1}$\ (}
\def\prv#1({Phys.\ Rev.\ $\us  {#1}$\ (}
\def\mpl#1({Mod.\ Phys.\ Let.\ $\us  {A#1}$\ (}
\def\ijmp#1({Int.\ J.\ Mod.\ Phys.\ $\us {A#1}$\ (}
\def\cqg#1({Class.\ Quantum Grav.\ $\us {#1}$\ (}
\def\anp#1({Ann.\ of Phys.\ $\us {#1}$\ (}
\def\tmp#1({Theor.\ Math.\ Phys.\ $\us {#1}$\ (}

\def\tit#1|{{\it #1},\ }
%
%%%%%%%%%%%%%%%%%%%%%%%%%%%%%%%%%%%%%%%%%%%%%%%%%%%%%%%%%%%%%%
%%%%% misc macros %%%%%
%%%%%%%%%%%%%%%%%%%%%%%%%%%%%%%%%%%%%%%%%%%%%%%%%%%%%%%%%%%%%%
\def\ni{\noindent}

\def\bar{\overline}
\def\us#1{\underline{#1}}

\def\hat{\widehat}
\def\to{\rightarrow}
\def\notin{\hbox{{$\in$}\kern-.51em\hbox{/}}}

\def\del{\partial}

 \def\ie{{\it i.e.}\ }
\catcode`\@=12

\def\cy{Calabi--Yau\ }
\def\K{K\"ahler\ }

%%%%%%%%%%%%%%%%%FRONTPAGE%%%%%%%%%%%%%%%%%%%%%%%%%%%%%%%%%%%%%%%%%%%

\def\aff#1#2{\centerline{$^{#1}${\it #2}}}
\hfill{CERN-TH.7384/94}

\hfill{POLFIS-TH. 07/94}
\def\pdate{August 1994}
\titlepage
\vskip .5truecm
\title
 {On the Geometry of  Moduli Space of Vacua in N=2 Supersymmetric
 Yang--Mills Theory
\foot{Supported in part by DOE
 grants DE-AC0381-ER50050 and DOE-AT03-88ER40384,Task E.}}
\vskip-.2cm
\author{
A.\ Ceresole$^{1}$, R.\ D'Auria$^{1}$ and
S.\ Ferrara$^{2}$}
\vskip2.truecm
\aff1{Dipartimento di Fisica, Politecnico di Torino,}
\centerline{\it  Corso Duca Degli Abruzzi 24, 10129 Torino, Italy}
\centerline{\it and}
\centerline{\it INFN, Sezione di Torino, Italy}
\line{\hfill}
\aff2{CERN, 1211 Geneva 23, Switzerland}
\line{\hfill}
%%%%%%%%%%%%%%%%%%%%%%%%%%%%%%%%%%%%%%%%%%%%%%%%%%%%%%%%%%%%%%%%%
\vskip-.8 cm
\def\abs
{\ni
We consider generic properties of the moduli space of vacua in $N=2$
supersymmetric Yang--Mills theory recently studied by Seiberg and Witten.
We find, on general grounds, Picard--Fuchs type of differential equations
expressing the existence of a flat holomorphic connection, which for
one parameter  (\ie for gauge group $G=SU(2)$), are  second order equations.
In the case of coupling to gravity ( as in string theory), where also
``gravitational'' electric and magnetic monopoles are present, the
electric--magnetic S duality, due to quantum corrections, does not seem any
longer to be related to $Sl(2,\IZ)$ as for $N=4$ supersymmetric theory.
}
\abstract{\abs}
\vfill
\endpage
\baselineskip=14pt plus 2pt minus 1pt
%%%%%%%%%%%%%%%%%%%%%%%%%%%%%%%%%%%%%%%%%%%%%%%%%%%%%%%%%%%%%%%%%%%%%%%%
%%%%%%%%%%%%REFERENCES%%%%%%%%%%%%%%%%%%%%%%%%%%%%%%%%%%%%%%%%%%%%%%%%%%
\ldf\seiwit{N.\ Seiberg and E.\ Witten, \tit Electric--Magnetic Duality,
Monopole Condensation, And Confinement In $N=2$ Supersymmetric Yang--Mills
Theory | preprint RU-94-52, IAS-94-43 , hep-th 9407087.}
\ldf\seib{N. Seiberg, \plt206B (1988) 75.}
\ldf\spec{
N. Seiberg, \nup303 (1988) 206;
A.\ Strominger, \cmp133 (1990) 163;
L.\ Castellani, R.\ D'Auria and S.\ Ferrara, \plt241B (1990)
57; \cqg1 (1990) 317;
P.\ Candelas and X.\ de la Ossa, \nup355 (1991) 455;
S.\ Ferrara and J.\ Louis, \plt278B (1992) 240;
A.\ Ceresole, R.\ D'Auria, S.\ Ferrara, W.\ Lerche and J. Louis,
\ijmp8 (1993) 79}
\ldf\essays{For a recent review, see A.\ Ceresole, R.\ D'Auria,
 S.\ Ferrara, W.\ Lerche,  J.\ Louis and T.\ Regge, \tit Picard--Fuchs
Equations, Special Geometry and Target Space Duality | preprint
CERN-TH.7055/93,
POLFIS-TH.09/93, to appear on \tit Essays on Mirror Manifolds | vol. II,
 S.\ T.\ Yau editor, International Press (1994).}
\ldf\fors{
A.\ C.\ Cadavid and S. Ferrara, \plt267B (1991) 193;
W.\ Lerche, D.\ Smit and N.\ Warner, \nup372 (1992) 87.}
\ldf\filq{A.\ Font, L. Ibanez, D.\ Lust, F.\ Quevedo, \plt249B (1990) 35.}
\ldf\bate{A.\ Erd\'elyi, F.\ Obershettinger, W.\ Magnus and F. G. Tricomi,
\tit Higher Trascendental Functions|, McGraw Hill, New York, (1953).}
\ldf\nfour{
A.\ Sen, \nup388 (1992) 457 and \plt303B (1993) 22;
A.\ Sen and J.\ H.\ Schwarz, \nup411 (1994) 35; \plt312B (1993) 105;
M.\ J.\ Duff and R.\ R.\ Khuri, \nup411 (1994) 473;
J.\ Gauntlett and J.\ A.\ Harvey, preprint EFI-94-36;
L.\ Girardello, A.\ Giveon. M.\ Porrati and A.\ Zaffaroni,
preprint NYU-TH-94/06/02.}
\ldf\tsd{For a review on target space duality see A.\ Giveon,
M.\ Porrati and E.\ Rabinovici, preprint RI-I-94, hep-th/9401139.}
\ldf\csf{E.\ Cremmer, J.\ Scherk and S.\ Ferrara, \plt074B (1978) 61.}
\ldf\susu{B.\ de Wit and A.\ van Proeyen, \nup245 (1984) 89;
E.\ Cremmer, C.\ Kounnas, A.\ van Proeyen, J.\ Derendinger, S.\ Ferrara,
B.\ De Wit and L.\ Girardello, \nup250 (1985) 385;
B.\ de Wit, P.\ G.\ Lauers and A.\ Van Proeyen, \nup255 (1985) 569;
S.\ Cecotti, S.\ Ferrara and L.\ Girardello, \ijmp4 (1989) 2475.}
\ldf\fklz{S.\ Ferrara, C.\ Kounnas, D.\ Lust and F.\ Zwirner, \nup365
 (1991) 431 .}
\ldf\cdgp{P.\ Candelas, X.\ de la Ossa, P.\ S.\ Green and L.\ Parkes,
\plt258B (1991) 118; \nup359 (1991) 21.}
%%%%%%%%%%%%%%%%%%%%%%%%%%%%%%%%%%%%%%%%%%%%%%%%%%%%%%%%%%%%%%%%%%%%%%%%
%%%%%%%%%%%%%%%%%%%%%%%%%%%%%%%BODY%%%%%%%%%%%%%%%%%%%%%%%%%%%%%%%%%%%%%
%%%%%%%%%%%%%%%%%%%%%%%%%%%%%%%%%%%%%%%%%%%%%%%%%%%%%%%%%%%%%%%%%%%%%%%%
\def\bB{{\bar B}}

Recently it has been shown that general properties of electric-magnetic
 duality, which is eventually linked to a conjectured dilaton-axion
 duality in superstring theories\filq , can be described in a fairly
general way in $N=4$\nfour\  and $N=2$\seiwit\
supersymmetric Yang-Mills theories.
However, $N=2$ Yang-Mills theories look much more interesting since both
perturbative and non perturbative phenomena, absent for $N=4$\seib\ , play
an important r\^ole in the discussion and determination of the electric--
magnetic duality. Due to the fact that the moduli space of $N=2$
Yang--Mills vacua is given by an $N=2$ K\"ahlerian space of a
particular kind\susu\ , it turns out that electric--magnetic duality is
described by a monodromy group $\Gamma$ which is a subgroup of
$Sp(2r,\IZ)$ where $r$ is the rank of the Yang--Mills group $G$.
For the case $r=1$ ($G=SU(2)$), Seiberg and Witten identified $\Gamma$
to be the group $\Gamma_2\subset Sp(2,\IZ)\simeq Sl(2,\IZ)$\seiwit .
The monodromies are reminiscent of a similar problem that arises in the
analysis of Calabi--Yau moduli space\spec\  where the monodromy group
 $\Gamma$,
related to the target space duality group\tsd\ , is a discrete subgroup
of $Sp(2 h_{21}+2,\IZ)$ and is related to the three--form cohomology.

In this paper we show that, as expected, the monodromy related to
electric--magnetic duality arises from ``Picard--Fuchs'' equations\essays\
which are associated to the rigid special geometry of $N=2$ supersymmetric
Yang--Mills theories. Following lines similar to those concerning  the
special geometry of Calabi--Yau moduli space\susu , we shall first give a
resum\'e of ``rigid special geometry'' in a coordinate free way and
then write the associated system of differential equations, which
always can be interpreted as the existence of a flat holomorphic
connection on a certain holomorphic bundle.

Let us first remind that if one has $n$ abelian vector multiplets
(here $n=r$ since we consider generic flat directions of a pure Yang--Mills
theory with gauge group $G$ broken to $U(1)^r$) their scalar fields,
in a supergravity basis, describe a K\"ahlerian sigma model
$$
G_{A\bB} \del X^A \del {\bar X}^B \eqn\uno
$$
with metric
$$
G_{A\bB}=-2\ {\rm Im}\ \del_A\del_\bB F=\del_A\del_\bB i
(F_C{\bar X}^C-{\bar F}_C X^C )\ \,
\eqn\due
$$
where $F$ is a holomorphic function of $X$ (${\bar \del}F=0$ ).
These coordinates are the analogue of the ``special coordinates'' in
the context of Calabi--Yau moduli space. A general coordinate free way
of describing the special geometry for that case was given in \susu\
and the associated system of Picard--Fuchs equations, together with
the flat holomorphic geometry, was discussed in \doubref\susu\essays\ .
The Riemann tensor of special geometry satisfies
$$
R_{\a{\bar \b}\c{\bar \d}}=g_{\a{\bar \b}}g_{\c{\bar \d}}+
g_{\c{\bar \b}}g_{\a{\bar\d}}
- e^{2K}W_{\a\b\e}W_{\bar \b \bar \d \bar \e} G^{\e\bar \e}\ ,
\eqn\trep
$$
where $G_{\e\bar \e}=\del_\e\del_{\bar \e}K$ is the K\"ahler metric and
$W_{\a\b\c}$ is a totally symmetric holomorphic tensor.

Here we consider rigid special geometry, where the moduli space is simply a
\K rather than a \K-Hodge manifold, and the constraint \trep\ becomes
$$
R_{\a{\bar \b}\c{\bar \d}}=-W_{\a\b\e}W_{\bar \b \bar \d \bar \e} G^{\e\bar \e}
\ .
\eqn\tre
$$
In the $X$ coordinates $G_{\e\bar \e}$ is given by
eq. \due\ and
$$
W_{ABC}=\del_A\del_B\del_C F\ . \eqn\quattro
$$
To promote formulae \due ,\ \quattro\  to arbitrary coordinates, one introduces
$n$ holomorphic functions $X^A(z)$ and a function $F(X^A(z))$. Then the
K\"ahler potential is
$$
K(z,\bar z)= i (F_A \bar X^A-\bar F_A X^A)\ \ \ (F_A={{\del F}\over{\del X^A}})
\ , \eqn\sei
$$
and
$$
\eqalign{
G_{\a\bar \b} &=\del _\a X^A\del_{\bar \b} \bar X^{B}
            {\del\over{\del X^A}}{\del \over {\del \bar X^B}} K\cr
W_{\a\b\c} &=\del_\a X^A\del_\b X^B\del_\c X^C\del_A\del_B\del_C F}\ .
\eqn\otto
$$
It is convenient, as in ref. \doubref\susu\essays  ,
 to introduce the flat vielbein
$$
e_\a^A=\del_\a X^A \eqn\nove
$$
which is a $n\times n$ matrix. The Christoffel connection $\G^{\c}_{\a\b}$
whose Riemann tensor satisfies
$$
\del_{\bar\c}\Gamma^\d_{\a\b}=G^{\d\bar\d}R_{\a\bar\c\b\bar\d}=-
W_{\a\b\epsilon}\bar W_{\bar\c\bar\d\bar\epsilon}G^{\epsilon\bar\epsilon}
G^{\d\bar\d} \eqn\dieci
$$
can be written as
$$
\Gamma^{\d}_{\a\b}(z,\bar z)=T^{\d}_{\a\b}(z,\bar z)+\hat\Gamma^{\d}_{\a\b} (z)
\ , \eqn\undici
$$
where
$$
\eqalign{
T^\d _{\a\b}(z,{\bar z}) &=e^A_\a e^B_\b \del_B G_{A\bar D}
 G^{-1\bar D C}e^{-1\d}_{\ \ C}\cr
\hat \Gamma^\d_{\a\b}(z) &= \del_\b e^A_\a e^{-1\d}_{\ \ A}\cr}\ .
\eqn\dodici
$$
{}From \dodici\ eq. \dieci\ immediately follows.
$\hat\G$ defines a flat connection in an $n\times n$ space: $R(\hat\G)=0$.

If one introduces the $2n$ objects $X^A(z), F_A(z)$ and the
$2n$ dimensional vector $V=(X^A,F_A)$, it is easy to show that the
following identities
hold
$$
\eqalign{
D_\a V &=V_\a\cr
D_\a V_\b &=- i\ W_{\a\b\c} V^\c\ \ \ \ \ V^\c=G^{\c\bar\c}\ \bar V_{\bar\c}\cr
D_\a\bar V_{\bar\c} &=0\cr}
\eqn\tredici
$$
where $D_\a$ is the covariant derivative in the original \K manifold.
To this non holomorphic system of identities it is associated an
holomorphic system, which is obtained by replacing $D_\a$ with the flat
covariant derivative $\hat D_\a$ where $\G\to\hat\G$, and $V^\c$ is
replaced by a holomorphic vector
$$
V^\a= (0,e^{-1\a}_{\ \ A})\ .\eqn\quattordici
$$
Since the first equation is left invariant by constant translations
$V\to V+c$, it is actually possible to consider $(V,V_\a,V^\a)$ as
$(2N+1)$ vectors so that
$$
\eqalign{
V &=(1,X^A,F_A)\cr
V_\a &=(0,e^A_\a, e^B_\a F_{AB})\cr
V^\a &= (0,0,e^{-1\a}_{\ \ A})\cr}
\eqn\quindici
$$
In terms of the $(2n+1)\times (2n+1)$ matrix
$$
\cV=\pmatrix{V \cr V_\b\cr V^\b \cr}\ ,
\eqn\sedici
$$
the holomorphic system can be written as
$$
\cD_\a \cV=0\ ,
\eqn\diciassette
$$
with
$$
\cD_\a=\del_\a-\cA_\a
\eqn\diciotto
$$
and the flat connection $\cA_\a$ given by
$$
\cA_\a=\pmatrix{ 0 &\delta_\a^\c &0\cr
                 0 &\hat\G_{\a\b}^\c & (W_\a)_{\b\c}\cr
                 0 & 0 & -\hat\G_{\a\c}^\b\cr}\ .
\eqn\diciannove
$$
Note that there is a $ISp(2n)$ acting on the right hand side of $V$
(or $\cV$) represented as $\pmatrix{1&0\cr C&M\cr
}$ with $M\subset Sp(2n)$. This follows from the fact that the
submatrix $\pmatrix{\hat\G^\c_{\a\b}
&(W_\a)_{\b\c}\cr0&-\hat\G^\b_{\a\c}\cr}$ is valued in the lie algebra of
$Sp(2n)$ (with respect to the metric $Q=\pmatrix{0&-1_n\cr
1_n&0\cr}$).

In special coordinates $e^A_\a=\delta^A_\a$,\ $\hat\G=0$ and the
connection $\cA_\a$ reduces to
$$\cA_\a =\pmatrix{0&\d^\c_\a&0\cr0&0&W_{\a\b\c}\cr0&0&0\cr}
\eqn\venti
$$
which is nilpotent of degree three ($(\cA_{\a})^3=0$).

Eq. \diciassette\ can be rewritten as a system of third order differential
equations for the upper component of $\cV$,
$$
\hat D_\a( W^{-1\hat\c})^{\e\b} \hat D_{\hat\c}\del_\b V=0\ ,
\eqn\diciab
$$
(where $\hat \c$ is a priori not summed over ). Actually,
since the first equation in \tredici\
can be used to just start with
$V_\a$, \diciab\  can be reduced to a second order
differential equation for $V_\b=\del_\b V$.
We can then delete the first entry in \quindici
$$
\eqalign{
V_\a &=(e^A_\a, e^B_\a F_{AB})\cr
V^\a &=(0,e^{-1\a}_A)\cr}
\eqn\ventiquattro
$$
and write the connection as $\cA_\a=\pmatrix{\hat\G^\c_{\a\b}
&(W_\a)_{\b\c}\cr0&-\hat\G^\b_{\a\c}\cr}$,
which reduces in special coordinates to
$$
\cA_\a=\pmatrix{0&(W_\a)_{\b\c}\cr0&0\cr}
\eqn\venticinque
$$
which is then nilpotent of degree two,
 and $W_{\a\b\c}$ is an $n$--dimensional
abelian subalgebra. The physical meaning of $W_{\a\b\c}$ is that they are
related to the Riemann tensor over the moduli space by eq. \tre.

In the case of one variable ($n=1$),  equation \diciab\ becomes
$$
(\hat D W^{-1} \hat D \hat \del)V=0\ ,
\eqn\ventuno
$$
and setting $U=\del V$ it becomes
$$
(\del +\hat\G) W^{-1}(\del-\hat\G)U=0\ .
\eqn\ventidue
$$
This yields a second order equation
$$
\del^2 U+a_1 \del U+a_0 U=0\ ,
\eqn\ventisei
$$
with
$$
\eqalign{
a_1 &=-\del \log W\cr
a_0 &=\del\log W \hat \G -\del\hat\G+\hat\G^2\ \ \ \ \ \hat\G=\del\log e \cr}
\eqn\vsette
$$
so that knowing $a_1, a_0$ one can compute $W$ and $e$
 (e=1 in special coordinates).
Note that the general solution in the one parameter case is
$$
U=\del V = (e, e {{\del^2 F}\over{\del X^2}})\ ,
\eqn\votto
$$
where $e$ is the vielbein component that here plays the r\^ole of a
rescaling factor. Taking the ratio of the two solutions one gets that
$\tau={{\del^2 F}\over{\del X^2}}$
 is the uniformizing variable for which the differential
equation reduces to ${d^2\over{d\tau^2}}(\ )=0$.
This is consistent with the fact that the metric of the
effective supergravity theory is
$$
G_{z\bar z}=|e(z)|^2 {\rm Im}\ \tau =|e(z)|^2 {\rm Im}\
 {{\del^2 F}\over{\del X^2}} >0
\eqn\vnove
$$
and therefore manifestly positive\seiwit .

As an explicit example, let us derive the differential equation \ventisei\
for the particular one parameter case  of Seiberg and Witten \seiwit.
Consider the family $E_u$ of genus one Riemann surfaces
$$
y^2=(x+1)(x-1)(x-u) \ ,
\eqn\wuno
$$
which, in homogeneous coordinates ($x\to{x\over{z}},y\to{y\over{z}}$)
can be described by the vanishing of the homogeneous polynomial $\cW$
in $CP^2$,
$$
\cW(x,y,z)=-z y^2+x(x^2-z^2)-u\ z(x^2-z^2)\ .
\eqn\wdue
$$
For convenience, we change  variables to ($x\to x+z,z\to x-z$), obtaining
$$
\cW=-(x-z) y^2+ x z (x-z)-u\ xz(x-z)\ .
\eqn\wtre
$$
The differential equation associated to \wtre\  can now be derived using
standard techniques, familiar from topological Landau--Ginzburg theories
\fors. Define the  integrals
$$
U_0 =\int {\omega\over{\cW}}\ \ \ , \ \ \ U_1\equiv{{dU_0}\over{du}}=
\int {\omega\over{\cW^2}}xz(x-z)
\eqn\wquat
$$
where $\omega$ is a volume form, which form a basis of the cohomology
$H^1(E_u)$.
By differentiating under the integral sign and using the ``vanishing
relations'' ${{\del\cW}\over{\del x}}={{\del\cW}\over{\del y}}=
{{\del\cW}\over{\del z}}=0$, one can show that
the vector $\pmatrix{U_0\cr U_1\cr}$ satisfies a regular, singular matrix
differential equation
$$
{d\over{du}}\pmatrix{U_0\cr U_1\cr}=\pmatrix{0&1\cr {{2u}\over{(1-u^2)}}&
{1\over{4(1-u^2)}}\cr}\pmatrix{U_0\cr U_1\cr}\ ,
\eqn\wsette
$$
which is equivalent to the second order differential equation
$$
{{d^2 U_0}\over{du^2}}-{{2u}\over{(1-u^2)}}{{d U_0}\over{du}}
-{1\over{4(1-u^2)}}U_0=0\ .
\eqn\wcin
$$
%Furthermore, eliminating the first order derivative in \wcin\ by the shift
%$U_0\to e^{{1\over2}\int{{2u}\over{1-u^2}}du} U_0$ one gets
%$$
%U_0''+{{3+u^2}\over{4(1-u^2)}}U=0
%\eqn\wotto
%$$
Comparing \wcin with \ventisei,\ \vsette\ one can read out that
$$
W={1\over{(u^2-1)}}
\eqn\yuka
$$
and that $\hat\Gamma$ satisfies
$$
-{{2u}\over{u^2-1}}\hat\Gamma-\del\hat\Gamma+\hat\Gamma^2={1\over{4(u^2-1)}}\ .
\eqn\conne
$$
Writing $\hat\Gamma=\del log\ e$, then \conne\ coincides with equation \wcin\
for $U_0$, in agreement with the fact that, according to \votto, $e$ is one
of the solutions of \wcin.

As a check, we may compute the asymptotic behaviour of the solutions
$(U_0^{(1)},U_0^{(2)})$ of this fuchsian equation around the singular points
$u=1,-1,\infty$. We find
$$
\eqalign{
u\to\pm 1\ \ \ \ (U_0^{(1)},U_0^{(2)}) &\approx   (c_1, log (u\mp 1))\cr
u\to\infty\ \ \ \ (U_0^{(1)},U_0^{(2)}) &\approx  (u^{-1/2},u^{-1/2}log u)\cr}
\ .\eqn\asuno
$$
Recalling that $U_0^{(1)}=\del V^{(1)},U_0^{(2)}=\del V^{(2)}$ and the
third period
$V^{(3)}=c$, one finds
$$
\eqalign{
u\to\pm 1\ \ \ \ (V^{(1)}, V^{(2)}+V^{(3)})&\approx(u\mp 1,c+c'(u\mp 1)
log(u\mp 1) )\cr
u\to\infty\ \ \ \ (V^{(1)},V^{(2)}+V^{(3)})&\approx(u^{1/2},u^{1/2}log u)\cr}
\eqn\asdue
$$
in agreement with the behaviour of the periods $(a,a_D)$ of \seiwit.

The change of variables $u\to 1-2z$, puts eq.\wcin\ into the form of
an hypergeometric equation of parameters $({1\over 2},{1\over 2},1)$
so that $U_0^1=\ _2\!F_1\left[{1\over2},{1\over2},1;{{1-u}\over{2}}\right]$.
Using the standard relations among hypergeometric functions \bate\ one
could also reconstruct the monodromy matrices of the periods in a
symplectic basis  as given in \seiwit .

The geometry of the moduli space is actually remarkably different when
gravitational degrees of freedom are introduced \nfour .
The reason is that in that case there are always two additional $U(1)$
factors, one coming from $G_{\mu i}$ and the other from $B_{\mu i}$.
One is the $N=2$ graviphoton and the other is the vector partner of
the dilaton--axion multiplet. Therefore one gets a $U(1)^{r+2}$
abelian algebra and at least $r+1$ vector multiplets. In string theory
with maximal $G$, $r=22$ and the special \K manifold has dimension
$r+1$. If the gauge group would be taken to be $SU(2)$ ($r=1$), then
the holomorphic prepotential would be of the form\doubref\seiwit\seib\
(in special coordinates $s= {{X_1}\over{X_0}}, t={{X_2}\over{X_0}}$,
and for instanton number n)
$$
F(s,t)=s t^2 + f_{one\ loop} (t)+\sum_{n=1}^\infty
 C_n\ t^2 ({{\Lambda^2}\over{t^2}})^{2n}e^{2\pi i n s}
\eqn\trenta
$$
with $s=i {{4\pi}\over{g^2}}+{{\theta}\over{2\pi}}$, where $f_{one\ loop}(t)$
does not violate the $s$ (dilaton--axion) Peccei--Quinn symmetry and
the non perturbative part gives the space time instanton contribution.
The generalization of this formula for $G=SU(N)$ is straightforward\seib .
Moreover, the metric of the moduli space will be that of special geometry
\doubref\spec\essays\ .

Unlike the rigid case discussed in \seiwit, it is natural to conjecture
here a monodromy in two variables and a central charge of the type
\doubref\seiwit\fklz\
$$
Z=\sum_{A=1}^{3} N^A_{(m)} F_A-M^A_{(e)}X_A
\eqn\tuno
$$
where $A=0,s,t$. The new $0,s$ components correspond to gravitationally
electrically and magnetically charged states.
This formula is analogous to the one suggested in \fklz\ for the massive
Kaluza-Klein and winding states for $(2,2)$ supersymmetric compactifications.
There, $(X^A,F_A)$ play the r\^ole of periods of holomorphic three-form
and duality is manifest with respect to monodromy in the moduli space of
 $(2,2)$ vacua\doubref\spec\essays .

In this case  the monodromy
group, \ie the duality group, would not be in $Sp(2,\IZ)$ but rather in
$Sp(6,\IZ)$ with a Picard--Fuchs system identical in form to the two--
parameter case of a \cy moduli space. In this case there is an
intriguing analogy between the moduli space of $N=2$ supersymmetric
Yang--Mills theory coupled to supergravity (with gauge group $G$ of
rank $r$) and \cy moduli space for the three-form cohomology with
hodge number $h_{21}=r+1$. The modular forms with respect to $\Gamma$
should reconstruct the full holomorphic function $F(s,t)$.
We further remark that the $Sl(2,\IZ)$ symmetry associated to
dilaton--axion (S) duality is peculiar of $N=4$ theories only, because
of the absence of quantum corrections. Indeed, for $N=2$ supersymmetric
 theories, the monodromy group associated to the periods
$ (1,s,t, F_s,F_t,2F-sF_s-tF_t)$ is expected to be a discrete group
$\Gamma\in Sp(6,\IZ)$, which will not in any way be related to
 $Sl(2,\IZ)$ or any of its subgroups. It is merely a property of the
tree-level uncorrected prepotential $F(s,t)=st^2$ to exhibit a non--
linear $Sl(2,\IR)$ symmetry (containing $Sl(2,\IZ)$) previously
found in $N=4$ supergravity\csf . This is similar to the example of
the one--parameter \cy moduli space whose metric for large volume
approaches the metric of ${{SU(1,1)}\over{U(1)}}$ homogeneous space\cdgp .

\ack{We would like to thank R. Stora for useful discussions.}

\refout
\end